\documentclass[reprint,superscriptaddress,prl,twocolumn,amsmath,amssymb,aps,longbibliography,hidelinks]{revtex4-1}

\usepackage{graphicx}
\usepackage{soul}
\usepackage{dcolumn}
\usepackage{bm}
\usepackage{hyperref}
\usepackage{xr-hyper}
\usepackage{xcolor}
\usepackage[normalem]{ulem}

\DeclareMathOperator{\csch}{csch}

\usepackage{mathtools}
\newcommand\underrel[3][]{\mathrel{\mathop{#3}\limits_{%
			\ifx c#1\relax\mathclap{#2}\else#2\fi}}}

\begin{document}

\title{Boosting macroscopic diffusion with local resetting}
	
	\author{Henry Alston}
	\affiliation{Department of Mathematics, Imperial College London, South Kensington, London SW7 2AZ, United Kingdom}
	\author{Thibault Bertrand}%
	\email{t.bertrand@imperial.ac.uk}
	\affiliation{Department of Mathematics, Imperial College London, South Kensington, London SW7 2AZ, United Kingdom}

\date{\today}

\begin{abstract}
\noindent Stochastic interactions generically enhance self-diffusivity in living and biological systems, e.g. optimizing navigation strategies and controlling material properties of cellular tissues and bacterial aggregates. Despite this, the physical mechanisms underlying this nonequilibrium behavior are poorly understood. Here, we introduce a model of interactions between an agent and its environment in the form of a local stochastic resetting mechanism, in which the agent's position is set to the \textit{nearest} of a predetermined array of sites with a fixed rate. We derive analytic results for the self-diffusion coefficient, showing explicitly that this mechanism enhances diffusivity. Strikingly, we show analytically that this enhancement is optimized by regular arrays of resetting sites. Altogether, our results ultimately provide the conditions for the optimization of the macroscopic transport properties of diffusive systems with local random binding interactions.
\end{abstract}

\maketitle


Dynamic interactions drive biological systems away from thermodynamic equilibrium \cite{Alston2022b}. Pili-mediated forces between bacteria such as \textit{Neisseria Meningitidis} \cite{Kuan2021, Ponisch2022}, where interactions are inherently stochastic due to the dynamic binding-unbinding of type-IV pili \cite{Bonazzi2018}, lead to the formation of nonequilibrium structures \cite{Alston2022a}. Biological agents with ligand-receptor contacts \cite{Bressloff2013, Marbach2022}, such as cells \cite{Ley2007}, protein cargos \cite{Allen2000} and viruses \cite{Muller2019}, employ similar cycles of attachment-detachment with their local environment to drive transport and self-assembly at much smaller lengthscales.

Diffusive motion interposed by random binding interactions to a confining surface can be studied in the wider context of so-called stochastic resetting, a paradigmatic model in nonequilibrium statistical physics \cite{Evans2011, Evans2020}. Resetting, where the particle position is moved instantaneously (``\textit{reset}'') to one of a predetermined set of sites, generically leads to a nonequilibrium steady-state \cite{Evans2011b, Eule2016, Mendez2016, Dibello2023} and finds applications in a broad range of simple yet biological relevant processes \cite{Boyer2014, Genthon2022, daSilva2022, Roldan2016, Reuveni2014, Basu2019, Karthika2020}. One striking example is the problem of optimizing target search strategies \cite{Condamin2007, Benichou2005, Majumdar2021, DeBruyne2022} where resetting events can lower the mean first-passage time to reach a target compared to simple diffusion across a range of lengthscales \cite{ Redner2001, Janson2012, Reuveni2016, Pal2017, Evans2013, Bray2013, Faisant2021, Besga2020,Benichou2009, Benichou2011}, from proteins binding to DNA to foraging animals.

Stochastic interactions generically enhance self-diffusion  in living matter. Pili-mediated interactions increase diffusivity at the surface of bacterial colonies \cite{Bonazzi2018}, imparting the aggregates with liquid-like properties \cite{Bonazzi2018, Kuan2021}; similarly, dynamically fluctuating cell-cell adhesion strengths lead to the fluidization of embryonic tissues \cite{Mongera2018}. Further, type-IV pili-mediated stochastic interactions can also drive persistent motion on surfaces through twitching motility \cite{Merz2000, Marathe2014, Simsek2019}. Binding-unbinding interactions between cargo and crosslinks in biopolymer networks, such as the extracellular matrix, can significantly increase the cargo's diffusivity by overcoming local trapping effects \cite{Goodrich2018}. Similar enhancement has been observed when a particle is confined in a channel with polymer chains grafted to each side: activity-driven attachment-detachment cycles can generate an increased effective diffusion of the target particle \cite{Lalitha2021}. Finally, enzyme-substrate chemical reactions have been argued to enhance the effective diffusivity of enzymes in solution \cite{Rezaei-Ghaleh2022, Golestanian2015, AgudoCanalejo2018a,AgudoCanalejo2018b,Jee2020}.

Though driven by diverse underlying processes, minimal physical models of stochastic interactions can establish generic constraints for realizing high-diffusivity behavior. In this Letter, we introduce a local (\textit{nearest-neighbor}) resetting mechanism, where the position of a diffusive particle is stochastically reset to the \textit{nearest} of a predetermined set of resetting sites. We propose this mechanism as a minimal implementation of local stochastic interactions such as attachment-cycles to fixed binding sites or intermittent attractive forces with static targets or neighboring agents. We confirm that resetting reduces the mean first-passage time to reach adjacent sites, then analytically exhibit an enhancement of the macroscopic self-diffusion coefficient of the tagged particle, offering a perspective that is generally overlooked for stochastic resetting models. Remarkably, we show explicitly that this enhancement is maximised by a regular array of resetting sites.

\begin{figure}[t!]
    \centering
    \includegraphics[scale=1]{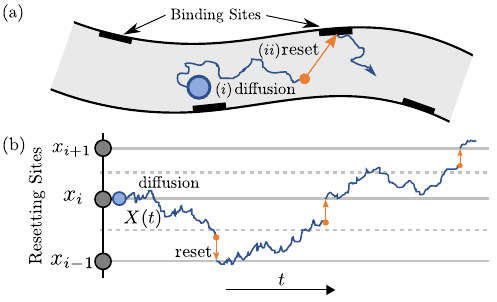}
    \caption{\textit{Diffusion with nearest-neighbor resetting ---} (a) Minimal model of diffusive agent subjected to stochastic binding and unbinding to its confining environment at static sites (e.g. modelling ligand-receptor contacts). (b) Example trajectory of a diffusive particle subjected to local resetting events. At each event, which are separated by exponentially distributed waiting times, the particle is moved instantaneously, \textit{reset}, to its nearest resetting site. 
    }
    \label{fig:schematic}
\end{figure}

\textit{Nearest-neighbor resetting mechanism. ---} Consider $N$ resetting sites $\{x_1, \dots x_{N}\}$ on the 1D domain $[0, L)$ with periodic boundary conditions, where without loss of generality, we assume that $x_1=0$. We will denote $\rho = N/L$ the density of resetting targets. We then consider the dynamics of a Brownian particle, whose motion is entirely characterized by the diffusion coefficient $D$, subjected to resetting events occuring with Poissonian rate $k_r$. At these events, the particle is moved to the nearest resetting site as shown in Fig.\,\ref{fig:schematic}.


In \cite{Note1}, we further motivate the instantaneous nature of the resetting mechanism through a formal separation of the diffusion- and interaction-timescales. While in what follows, we study analytically the case of instantaneous resetting events, we also probe the parameter space beyond this timescale separation (e.g. when resetting events have finite duration) and confirm that our results stand for a range of physically realizable parameters (see \footnote{See Supplemental Material at [] for further analytical and computational details, which includes Refs. []} for an extended discussion).

\begin{figure*}[t!]
    \centering
    \includegraphics[width=1\textwidth]{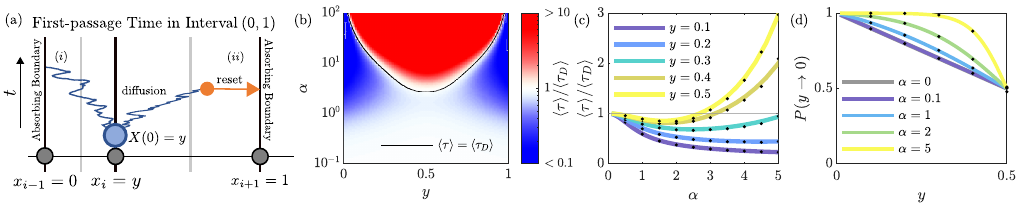}
    \caption{\textit{Dynamics of nearest-neighbor resetting on the unit interval ---} (a) We consider three adjacent resetting sites placed at $\{0, y, 1\}$, respectively. Initializing the particle at site $y$, we measure the first-passage time to reach either of the two adjacent sites, via (i) pure diffusion or (ii) a resetting event. (b) We compare the mean first-passage time for the resetting process against that of pure diffusion. (c) Comparing the two mean first passage times for different locations of the $y$ site. For small $\alpha = \sqrt{k_r \ell^2 / 4D}$, the resetting mechanism always wins, but this disappears at large $\alpha$ with the crossover value depending on $y$ as shown in (b). (d) The splitting probability to hit the $x=0$ boundary before the $x=1$ boundary. In (c) and (d), our results are compared to numerical simulations of the microscopic process (see details in \cite{Note1}).
    }
    \label{fig:interval}
\end{figure*}

\textit{Effective diffusion coefficient. ---} To assess the role of the resetting mechanism on the particle's dynamics, we study its long-time effective diffusion coefficent. To do so, we formally map our model onto a continuous-time random walk problem. Without loss of generality, we can consider that at time $t$ our particle is located at resetting site $x_n$. Working on the interval $[x_{n-1}, x_{n+1}]$, we need to study the statistics of the jumps $x_n \rightarrow x_{n+1}$ and $x_n \rightarrow x_{n-1}$, i.e. an escape from site $n$ to one of its nearest-neighbor resetting sites. This can happen either by {\it diffusion and resetting} or by {\it diffusion alone}. 

For each site $x_n$, we write the increase in the variance of the particle position due to the jump to one of the neighboring sites as 
\begin{align}
    \langle (\Delta x)^2\rangle_n&= (x_{n+1}-x_n)^2 \mathbb{P}(n\rightarrow n+1)\nonumber\\
    &\quad+ (x_{n}-x_{n-1})^2 \mathbb{P}(n\rightarrow n-1).
\end{align}
where $\mathbb{P}(n\rightarrow n\pm 1)$ are the splitting probabilities corresponding to the two possible events (reaching either end of the interval). 

Due to the Markovian nature of the process, we can write the mean-squared displacement for this continuous-time random walk \cite{Hughes1995} as 
\begin{equation}
    \langle (\Delta x)^2\rangle (t) = \sum_{n=1}^N \Lambda_n(t) \langle (\Delta x)^2\rangle_n
\end{equation}
where $\Lambda_n$ is the number of times that site $n$ has been escaped from up to time $t$. The effective diffusion coefficient can then be derived as
\begin{equation}\label{eq:eff_D}
    D_{\rm eff} = \lim_{t\to \infty}\frac{\langle (\Delta x)^2\rangle}{2t} = \sum_{n=1}^N \pi_n\frac{\langle (\Delta x)^2\rangle_n}{2\langle \tau\rangle _n}
\end{equation}
where $\pi_n$ is the fraction of the total trajectory time spent escaping from each site: $\pi_n = \lim_{t\to\infty}{\Lambda_n(t) \langle \tau\rangle_n}/{t}$ and $\langle \tau \rangle_n$ is the (unconditional) mean first-passage time (MFPT) to reach either of the nearest neighbors of site $n$. Eq.\,\eqref{eq:eff_D} can also be read as a time average of the local diffusion coefficients $D_{\rm eff}^{(n)}={\langle (\Delta x)^2\rangle_n}/2{\langle \tau\rangle_n}$. To make further analytical progress, we thus need to consider the local dynamics, namely the MFPT $\langle \tau \rangle_n$ and the splitting probabilities $\mathbb{P}(n\rightarrow n \pm 1)$.

\textit{First-passage time to adjacent sites. ---} We will first study the first-passage time distribution from site $x_n$ to adjacent resetting sites $\{x_{n-1},x_{n+1}\}$. Without loss of generality, we can translate this interval so that $x_n = 0$, denoting $x_{n-1}=-a < 0$ and $x_{n+1} = b>0$, respectively. We define $Q(t)$ as the probability that a particle has \textit{not} reached either site before time $t$. We write this survival probability as a sum over the different realizations of the resetting process: 
\begin{equation}\label{eq:qt_def}
    Q(t) = \sum_{m=0}^\infty \mathcal{P}(m,t) Q_m(t)
\end{equation}
where $\mathcal{P}(m, t)$ is the probability of observing exactly $m$ resetting events before time $t$ given by the Poisson distribution
\begin{equation}
    \mathcal{P}(m, t) = \frac{(k_r t)^me^{-k_r t}}{m!}
    \label{eq:poisson}
\end{equation}
and $Q_m(t)$ is the probability of a particle surviving to time $t$ given that there have been $m$ resetting events, all of which have reset the particle to $x_n$ rather than its adjacent sites.

For a particle to survive to time $t$, its position must satisfy $x(t')\in(-a, b)$ for $t'<t$, plus $x(\tau_{m'})\in(-a/2, b/2)$ at all resetting times $\tau_{m'}$ to ensure resetting to the site $x_n = 0$. If we suppose that there were $m$ resetting events before time $t$ occuring at $\{\tau_1, \dots, \tau_m\}$ (where $\{\tau_{m'}\}$ is a strictly increasing sequence), the conditional survival probability is
\begin{align}
    Q_m(t; \{\tau_{m'}\}) = P_s(\tau_1) &P_s(\tau_2-\tau_1)\cdots \nonumber \\
    		&P_s(\tau_m-\tau_{m-1})Q_0(t-\tau_m) 
    \label{eq:qmt}
\end{align}
where we have defined $P_s(\tau)$, the probability that the particle survives a resetting event at time $\tau$ given that it was initially at $x=0$ and $Q_0(\tau)$, the probability that a diffusive particle initially at $x=0$ does not escape the interval $(-a, b)$ by a time $\tau$. 

These probabilities can be evaluated using the solution $u(x, t)$ to the diffusion equation in one spatial dimension, with absorbing boundary conditions $u(-a, t)=u(b, t)=0$ and initial condition $u(x, t) = \delta(x).$ The probability distributions then required for evaluating Eq.\,(\ref{eq:qmt}) can then be defined through
\begin{equation}\label{eq:psnqs}
    P_s(\tau)=\int_{-a/2}^{b/2}dx \,u(x, \tau)~~\text{and}~~Q_0(\tau)\int_{-a}^b dx\, u(x, \tau),
\end{equation}
where $u(x, t)$ can be obtained through separation of variables \cite{Note1}. 

Armed with these, we can express $Q_m(t)$ as an average over all possible configurations of the resetting times sequence. For this, we argue that the probability of realizing a given set of resetting times $\{\tau_{m'}\}$ given that $m$ resetting events occur in a time $t$ is exactly given by $m!/t^m$; indeed, it is the probability of sampling $m$ uniformly-distributed times between $0$ and $t$ in any order. It follows that 
\begin{equation}\label{eq:qm_def}
    Q_m(t) = \frac{m!}{t^m}\int_0^{\tau_2}d\tau_1\dots \int_0^td\tau_m\: Q_m(t;\{\tau_{m'}\}).
\end{equation}

Defining here the Laplace transform of the survival probability as $\tilde{Q}(s) = \int_0^\infty dt \:e^{-st} Q(t)$, we can rewrite Eq.\,(\ref{eq:qt_def}) in Laplace space as 
\begin{align}
    \tilde{Q}(s) &= \sum_{m=0}^\infty (k_r \tilde{P}_s(s+k_r))^m\tilde{Q}_0(s+k_r) \\
    			&=\frac{\tilde{Q}_0(s+k_r)}{1-k_r\tilde{P}_s(s+k_r)}
\end{align}
from which one can derive the unconditional mean first-passage time to reach one of the nearest-neighbor resetting sites as 
\begin{equation}
    \langle \tau \rangle = \tilde{Q}(s=0) =  \frac{\tilde{Q}_0(k_r)}{1-k_r\tilde{P}_s(k_r)}.
\end{equation}

As shown in \cite{Note1}, we can find a general expression for the first-passage time density in Laplace space using Eqs.\,(\ref{eq:psnqs}). While not in general tractable, this expression takes a particularly simple form in the case where $\ell \equiv a = b$. In this case, the first-passage time density $\tilde{\tau}(s) \equiv 1-s\tilde{Q}(s) $ reads 
\begin{align}
    \tilde{\tau}(s)= \frac{s + k_r \cosh\bigg(\sqrt{\frac{(s+k_r)l^2}{4 D}}\bigg)}{k_r\cosh\bigg(\sqrt{\frac{(s+k_r)l^2}{4 D}}\bigg)+ s\cosh\bigg(2\sqrt{\frac{(s+k_r)l^2}{4 D}}\bigg)},
\end{align}
leading to the following succinct expression for the mean first-passage time 
\begin{equation}\label{eq:mfpt_lat}
    \langle \tau \rangle  = \frac{2}{k_r}\sinh(\alpha)\tanh(\alpha).
\end{equation}
where we have defined the dimensionless parameter $\alpha=\sqrt{k_r \ell^2 / 4D}$. This can be understood as the ratio of the two competing timescales in the problem: the time between two resetting events, $1/k_r$, and the typical time to diffuse outside of the resetting neighborhood of the initial site, which scales as $(\ell/2)^2/D=1/(4\rho^2D)$.

Through a suitable re-scaling of the interval $(-a,b)$, we consider for simplicity and without loss of generality the dynamics on the interval $(0, 1)$ with a site at intermediate point $y = a /(a+b)$ [see Fig.\,\ref{fig:interval}(a)] and identify in Fig.\,\ref{fig:interval}(b) that the resetting mechanism can lower the first passage time for a wide range of parameter values but can be a hindrance when $k_r \gg \rho^2 D $, i.e. when the timescale between resetting events is very short compared to the timescale for the particle to diffuse over length $1/\rho$. We also note that the reduction in mean first-passage time compared to pure diffusion is most notable when the initial site is near the boundaries, as shown in Fig.\,\ref{fig:interval}(c).

\textit{Splitting probabilities to adjacent sites. ---} To evaluate the splitting probabilities, we note that a particle starting in $x=0$ at time $t=0$ can reach an adjacent sites by either diffusion or a resetting event. The probability that the particle diffuses through one of the boundaries before a resetting event, denoted $\mathbb{P}(0\leadsto s)$ with $s \in \{-a,b\}$, is evaluated as the integrated-current through the boundary up until the initial resetting time which is exponentially distributed leading to
\begin{subequations}
\begin{align}
    \mathbb{P}(0 \leadsto -a) &= D\int_0^\infty dt\, k_r e^{-k_r t}\int_0^t d\tau\,\partial_x u \big|_{x=-a}, \\
    \mathbb{P}(0 \leadsto b) &= -D\int_0^\infty dt\, k_r e^{-k_r t}\int_0^t d\tau\,\partial_x u \big|_{x=b}.
\end{align}
\end{subequations}
Next, the probability that the tagged particle is reset to one of the adjacent sites, denoted $\mathbb{P}(0\mapsto s)$, is exactly
\begin{subequations}
\begin{align}
    \mathbb{P}(0\mapsto -a) &= \int_0^\infty dt \,k_r e^{-k_r t}  \int_{-a}^{-a/2}dx \,u(x, t), \\
    \mathbb{P}(0\mapsto b)  &= \int_0^\infty dt \,k_r e^{-k_r t}  \int_{b/2}^{b}dx \,u(x, t).
\end{align}
\end{subequations}
Finally, we can evaluate the splitting probabilities to reach each adjacent site $s \in \{-a,b\}$ by either diffusion or resetting as 
\begin{equation}
	\mathbb{P}(0\rightarrow s) = \frac{\mathbb{P}(0\leadsto s) + \mathbb{P}(0\mapsto s)}{\sum_{r\in \{-a, b\}}\mathbb{P}(0\leadsto r) + \mathbb{P}(0\mapsto r)}.
\end{equation}

Given $u(x,t)$ the solution to the above one-dimensional diffusion problem, we can evaluate these splitting probabilities \cite{Note1} as shown in Fig.\,\ref{fig:interval}(d). At small values of $\alpha$, the splitting probabilities are linear, as in the case of pure diffusion. This disappears at large $\alpha$, where we observe non-linear scaling and a strong preference to reach the nearest site.

\begin{figure}[t!]
    \centering
    \includegraphics[width=0.47\textwidth]{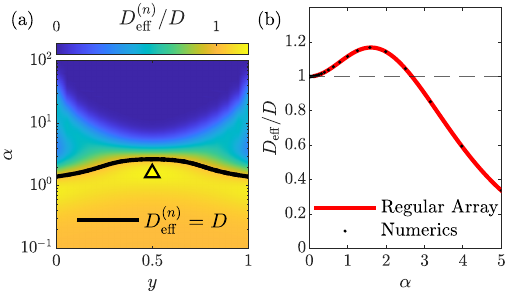}
    \caption{\textit{Effective diffusion coefficient on unit interval ---} (a) Local effective diffusion coefficient constructed from the splitting probabilities and unconditional mean first-passage time for the unit interval with resetting sites at $x = \{0, y, 1\}$, showing a boost in local diffusivity over a wide range of parameters and a global maximum (yellow triangle). (b) Macroscopic diffusivity as defined in Eq.\,\ref{eq:eff_D} is maximized for equidistant adjacent resetting sites (i.e for $y=0.5$) and $\alpha^* \approx 1.606$ where it takes the value $D_{\rm eff} \approx 1.169D$. }
    \label{fig:local_deff}
\end{figure}

\textit{Macroscopic diffusion coefficient for a regular array of resetting sites. ---} Armed with these results, we evaluate the local diffusion coefficients $D^{(n)}_{\rm eff}$ for all possible configurations of three adjacent sites. As before, we consider the dynamics on the unit interval and use the fact that the results can be extended to an arbitrary interval $\{-a, 0, b\}$ after a suitable re-scaling of space. As shown in Fig.\,\ref{fig:local_deff}(a), $D^{(n)}_{\rm eff}$ can be evaluated numerically from the analytic expressions above for any values of $\alpha$ and $y\in(0, 1)$ to show the existence of a global maximum at $\alpha\approx 1.606$ and $y=0.5$. 

We now turn to the macroscopic diffusion coefficient for a regular array of resetting sites. Obtaining closed form analytic results for arbitrary arrays of resetting sites requires the evaluation of the stationary probabilities $\bm{\pi}=[\pi_n]$, which in general is not tractable analytically. Nevertheless, for the case of a regular array of $N$ resetting sites (with spacing $\ell=1/\rho$) we have $\pi_n\equiv1/N$ and hence the large-scale effective diffusion coefficient can be expressed as 
\begin{equation}
D_{\rm eff}/D = \alpha^2\csch(\alpha)\coth(\alpha),
\end{equation}
which we plot in Fig.\,\ref{fig:local_deff}(b).

Strikingly, this implies that the macroscopic diffusion coefficient can indeed exceed the particle's bare diffusion coefficient when $\alpha < 2.676$ and is maximised for $\alpha^* \approx 1.606$ for which $D_{\rm eff}(\alpha^*) \approx 1.169 D$. This implies that this enhancement of diffusive properties is a {\it generic feature} of this model, going beyond the example of a regular array of resetting sites, provided the resetting occurs sufficiently rarely (see \cite{Note1} for an extended discussion). We conclude that our local resetting mechanism can induce a striking enhancement of the transport properties of a simple diffusive particle over macroscopic scales which we capture in an analytically tractable model of stochastic interactions.

\textit{Regular arrays of resetting sites optimize diffusion enhancement. ---} Finally, we study the conditions for optimization of the effective macroscopic diffusivity of the particle. Assuming that the resetting rate is an external parameter which we can set to maximize the self-diffusivity, we can write the optimal diffusion coefficient for an array of resetting sites $\{x_n\}_{n\in[1,N]}$ as
\begin{equation}\label{eq:deff_sum}
    \max_{k_r}\big(D_{\rm eff}/D) = \max_{k_r}\left(\sum_{n=1}^N  \pi_n \frac{\langle (\Delta x)^2\rangle_n}{2D\langle\tau\rangle_n}\right).
\end{equation}
The right-hand-side can be bounded from above trivially by the optimal value for the local effective diffusion coefficient as
\begin{equation}
    \max_{k_r}\bigg(\sum_{n=1}^N \pi_n \frac{\langle (\Delta x)^2\rangle_n}{2D \langle\tau\rangle_n}\bigg) \leq \max(D^{(n)}_{\rm{eff}}/D).
\end{equation}

However, the results of Fig.\,\ref{fig:local_deff}(a) imply that the maximum realizable value of $D^{(n)}_{\rm eff}/D$ is exactly the maximum achieved for a regular array of resetting sites leading to:
\begin{equation}\label{eq:max_deff}
    \max_{k_r}(D_{\rm eff}/D) \leq (\alpha^*)^2\csch(\alpha^*)\coth(\alpha^*)\approx 1.169.
\end{equation}
In other words, we have showed analytically that the maximal value for the effective macroscopic diffusivity of our process is obtained for regular arrays of resetting sites, giving in turn an upper-bound for the macroscopic effective diffusion coefficient over any array of resetting sites. 

We confirm this result two ways. First, we measure the diffusion coefficient for randomly generated arrays of resetting sites $\{x_n\}_{n\in [1,N]}$; to do so, we numerically compute the stationary probability vector $\bm{\pi}=[\pi_n]$ which allows us to calculate the maximum enhancement in diffusivity achievable through Eq.\,\eqref{eq:deff_sum}. We confirm that a regular array outperforms random arrays and show that increasing the variance in the local density of resetting sites decreases the diffusion enhancement. Second, we show that an optimization algorithm aiming at maximizing the effective diffusion coefficient of the particle over the arrangement of resetting sites starting from random configurations always converges to evenly spaced resetting sites \cite{Note1}. 

\textit{Discussion \& outlook. ---} We have introduced a local resetting mechanism as a minimal implementation of fluctuating interactions between a tagged particle and its local environment. Strikingly, we show that this local mechanism can generically enhance the macroscopic self-diffusion coefficient of the tagged particle for generic arrangements of resetting sites. We further show that this is maximized for regular arrays of evenly-spaced resetting sites. This implies that physical implementations of attachment-detachment cycles between a tagged particle and e.g. binding sites on a surface or static neighboring agents are optimized for enhancing self-diffusion when the sites are placed in a ordered manner, with minimal variance in the distances between neighboring sites. Our formal treatment of interactions with individual binding sites goes beyond previous coarse-grained approaches \cite{Marbach2022} and captures explicitly the impact of stochastic interactions on the effective diffusion coefficient. It also sufficiently captures the enhancement in diffusion observed in a range of biological settings \cite{Bonazzi2018, Mongera2018, Goodrich2018, Lalitha2021, Merz2000, Marathe2014, Simsek2019}. 

\begin{acknowledgments}
HA was supported by a Roth PhD scholarship funded by the Department of Mathematics at Imperial College London.
\end{acknowledgments}


%
	
\end{document}


\preprint{APS/123-QED}

\title{Supplemental Material for ``Boosting macroscopic diffusion with local resetting"}
	
	\author{Henry Alston}
 	\affiliation{Department of Mathematics, Imperial College London, South Kensington, London SW7 2AZ, United Kingdom}
	
	\author{Thibault Bertrand}%
	\email{t.bertrand@imperial.ac.uk}
	\affiliation{Department of Mathematics, Imperial College London, South Kensington, London SW7 2AZ, United Kingdom}

\date{\today}

\maketitle

\hrule

\tableofcontents

\vspace{2em}
\hrule

\section{Realistic resetting implementation}

In the main text, we assume the dynamics of interactions between the tagged particle and the array of sites to be instantaneous. In particular, during attachment, the tagged particle is moved instantaneously to the binding site and the particle detaches from the binding site instantaneously as well. This idealized and analytically tractable scenario is chosen in order to first confirm that a local binding-unbinding mechanism to a surface can indeed enhance self-diffusion, as shown explicitly in the main text. In this section, we motivate via a separation of timescales our model as the limit of a general binding-unbinding mechanism. We then show that our results remain qualitatively valid in this more general case for a wide range of parameters.

\subsection{Motivating nearest-neighbor resetting through a formal separation of timescales}

\begin{figure}
    \centering
    \includegraphics[scale=1.2]{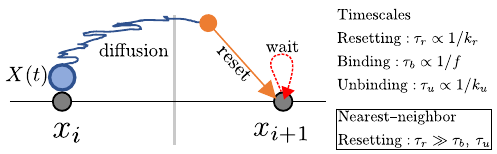}
    \caption{\textit{Schematic for full diffusive process with stochastic binding-unbinding mechanism---} We identify three relevant timescales in physical realizations of the stochastic process considered in the current work: (i) the time between contacts with the surface, labelled as the \textit{resetting} timescale here, (ii) the time it takes from initiation of the contact to pull the particle to the surface, here denoted the \textit{binding} timescale, and (iii) the characteristic time spent at the resetting site after arrival, i.e. the residence time at the site, denoted here \textit{unbinding} timescale. The main text results correspond to the formal timescale separation $\tau_r\gg \tau_b, \tau_u$.}
    \label{fig:schematic_full}
\end{figure}

As shown in Fig.\,\ref{fig:schematic_full}, a diffusive process with binding-unbinding dynamics to a surface can be decomposed into three distinct stages. Starting from the moment the particle is released from a binding site, the first stage corresponds to the time spent diffusing until the next binding interaction is initiated. In our model, this is precisely the timescale $\tau_r = 1/k_r$ as our resetting events are assumed to happen with Poissonian rate $k_r$. During the second stage, the particle is physically moved  to the binding site. If one assumes that this happens with constant force $f$, then the timescale associated with this step is $\tau_b \propto 1/f$ for an overdamped dynamics.  Finally, the particle remains at the binding site until it is released with constant rate $k_u$, leading to a characteristic residence time $\tau_u\propto 1/k_u$ at the binding site. 

If the pulling force dragging the particle to the binding site is large and the particle is released from the site soon after it arrives --- i.e. if  $\tau_r\gg \tau_b, \tau_u$ --- then the only relevant timescale in the problem is that between successive binding (or resetting) events, justifying the nearest-neighbor resetting model studied in the current work.

\subsection{Waiting-time distribution at resetting sites}
Now we suppose that there are two physically relevant timescales in the particle's dynamics: the resetting timescale $\tau_r$ and the unbinding timescale $\tau_u$ --- corresponding to the time spent at the binding site. Physically, this is realizable when $\tau_r, \tau_u\gg \tau_b$, i.e. in systems with strong interactions, such as the pili-mediated interactions between bacteria such as {\it Neisseria Meningitidis} \cite{Ponisch2022, Bonazzi2018}. 

In this case, the particle now spends a non-negligible fraction of its time stationary at each site before being released. If we suppose that the particle detaches from the binding site after resetting with some constant rate $k_u$ (i.e. the particle spends an exponentially-distributed length of time at the site after each resetting event) then the effective diffusion coefficient can be written in terms of $D_{\rm eff}^{\infty}$, the diffusion coefficient for $k_u=\infty$ (studied in the main text) as 
\begin{equation}\label{eq:diff_wtd}
    D_{\rm eff}(k_u<\infty)= D_{\rm eff}^{\infty} \bigg(\frac{k_u}{k_r + k_u}\bigg).
\end{equation}
Indeed, the diffusive dynamics are the same as in the $k_u=\infty$ case, but now the total trajectory time is increased due to a non-negligible time spent at the resetting sites. The factor $\frac{k_u}{k_r + k_u}$ is exactly the average fraction of the total trajectory time spent in the diffusive phase, i.e. not bounded to a site. When this rescaled trajectory time is accounted for in the expression for the diffusion coefficient, we arrive at Eq.\,\eqref{eq:diff_wtd}.

For the case of a regular array with sites evenly spaced by distance $\ell$, we can define $\tilde{k} = k_r/k_u$ and write the rescaled effective diffusion coefficient exactly as 
\begin{equation}
    D_{\rm eff}(k_u<\infty) / D = \alpha^2\csch(\alpha)\coth(\alpha)\bigg(\frac{1}{1 + \tilde{k}}\bigg).
\end{equation}
If $\tilde{k}<0.169\dots$, then we can still observe an enhancement in diffusion for some range of $\ell$, the distance between sites in the regular array. 

While the waiting time at sites in the current model represent a significant hindrence to self-diffusivity, binding-interactions themselves can be also be useful in re-arranging the local environment, such as in the case of cargo-crosslinker interactions in polymer gels \cite{Goodrich2018}. Here, interactions can open up pathways by transiently disrupting the local arrangement of the network and prevent self-trapping in the gel. In the current work, we overlook the possibility of such functionalities of the binding process in order to construct a most general model. 

\subsection{Finite-speed resetting mechnanism}
Finally, we assume that the resetting mechanism itself happens over a physically relevant timescale (corresponding to a resetting mechanism with finite speed). This must also be factored into our calculation of the diffusion coefficient. To isolate its impact on the calculated quantities, we suppose that the resetting timescale $\tau_r$ and the binding time $\tau_b$, are the only physically relevant timescales. We model this formally by assuming that the particle is reset by a pulling force of constant strength, $f = \gamma v$. For a regular array of binding sites, we can evaluate the average time it takes to reset the particle to its nearest site $\langle \tau_f \rangle(v)$ by evaluating the average distance between particle and nearest site when resetting occurs. This is exactly given by 
\begin{equation}\label{eq:t_ni_reset}
    \langle \tau_f\rangle = \frac{1}{v}\int_0^\infty dt\frac{k_r e^{-k_r t}}{\sqrt{4\pi D t}}\int_{-\infty}^\infty dx\, e^{-\frac{x^2}{4Dt}}S(x).
\end{equation}
where $S(x)$ is the distance to the nearest resetting point, averaged over the distribution of the particle's position at the subsequent resetting event. We consider again the case of the lattice array of sites, for which $S(x)$ takes the shape of a triangle wave. We then evaluate Eq.\,\eqref{eq:t_ni_reset} as
\begin{equation}
    \langle \tau_f\rangle = \frac{\alpha\tanh(\alpha/4)}{k_rv_0}
\end{equation}
where we have defined the dimensionless quantity $v_0 = v\ell/D$ as the ratio of two timescales: the advection timescale $\ell/v$ and the diffusive timescale $\ell^2/D$.

The effective diffusion coefficient is then rescaled from its value for the instantaneous case by a factor $k_r^{-1}/(k_r^{-1} + \langle \tau_f\rangle)$, thus
\begin{equation}
    D_{\rm eff}(v < \infty) = D_{\rm eff}^{\infty}\frac{v_0}{v_0 + \alpha \tanh(\alpha/4)}.
\end{equation}

\begin{figure}
    \centering
    \includegraphics{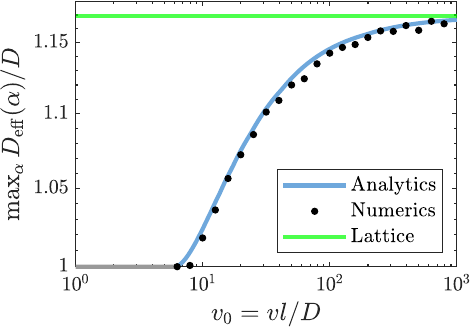}
    \caption{\textit{Effective diffusion coefficient for finite-speed resetting ---} We investigate to what extent the enhancement in diffusion is due to the instantaneous (and somewhat unphysical) nature of the resetting process. When the resetting happens at a constant finite speed $v$, the effective diffusion coefficient is reduced, but an enhancement is still observed provided $v_0 > 10$. Otherwise, the diffusion is maximized when there is no resetting at all.}
    \label{fig:nireset}
\end{figure}

This can be argued in the following way: the trajectory in space is indistinguishable from that of the infinite-speed resetting case, the difference is that it now happens over longer time (due to the non-zero fraction of time now spent in the resetting phase). Therefore, the diffusion coefficient can be written as the result for the infinite-speed resetting case, rescaled by a factor that accounts for this longer trajectory time which we identify as $k_r^{-1}/(k_r^{-1} + \langle \tau_f\rangle)$. This factor captures the fraction of time spent in the diffusive phase (as opposed to the resetting phase). For $v_0\gg 1$, the dynamics remain similar to the instantaneous case. As shown in Fig.\,\ref{fig:nireset}, where we compare analytics to numerical simulations of the process with finite-speed resetting, we observe a crossover point when $v_0\approx 10$ below which no enhancement in diffusion is observed.

\section{Dynamics in interval: First-passage time and splitting probability}

\subsection{First-passage time distribution}

First we consider the unconditional first-passage time distribution for a particle initialized at a resetting site at $0$ to reach either of the two neighboring sites located at $-a<0$ and $b>0$. We calculate the relevant properties by first solving for the spatial distribution of particles originally in $x=0$ that have not reached one of the boundaries at time $t$, assuming there have been no resetting events. This distribution solves $\partial_t u = D\partial_x^2 u$ with absorbing boundary conditions $u(-a, t) = u(b, t) = 0$ and initial condition $u(x, 0) = \delta(x).$ This problem can be solved analytically (e.g. via the separation of variables method):
\begin{equation}\label{eq:si_uxt}
    u(x, t) = \frac{2}{a+b}\sum_{n=1}^\infty \sin\left(\frac{n \pi a}{a+b}\right) \sin\left(\frac{n\pi(a+x)}{a+b}\right) \exp\left(-\frac{D n^2 \pi^2t}{(a+b)^2}\right).
\end{equation}

For the first passage time distribution, we recall the expression from the main text:
\begin{equation}\label{eq:SI_fptd}
    \tilde{\tau}(s) = 1-s\tilde{Q}(s) = 1-\frac{s\tilde{Q}_0(s+k_r)}{1-k_r\tilde{P}_s(s+k_r)}
\end{equation}
where the mean first-passage time is given by
\begin{equation}\label{eq:SI_mfpt}
    \langle \tau\rangle = \tilde{Q}(s=0) =  \frac{\tilde{Q}_0(k_r)}{1-k_r\tilde{P}_s(k_r)}.
\end{equation}
To derive analytic expressions, we return to the definitions for $Q_0$ and $P_s$:
\begin{equation}\label{eq:psnqs}
    P_s(\tau)=\int_{-a/2}^{b/2}dx \,{u}(x, \tau)\quad\text{and}\quad Q_0(\tau)=\int_{-a}^b dx\, {u}(x, \tau),
\end{equation}
which we evaluate using Eq.\,\eqref{eq:si_uxt} as 
\begin{equation}
    P_s(\tau) = \sum_{n=1}^\infty\frac{2}{\pi n}  \sin \left(\frac{\pi  a n}{a+b}\right) \left(\cos \left(\frac{\pi  a n}{2 (a+b)}\right)-\cos \left(\frac{\pi  n (2 a+b)}{2 (a+b)}\right)\right) \exp\left(-\frac{D n^2 \pi^2t}{(a+b)^2}\right)
\end{equation}
and
\begin{equation}
    Q_0(\tau) = \sum_{n=1}^\infty \frac{2 (1-(-1)^n)}{\pi n} \sin\left(\frac{\pi  a n}{a+b}\right)\exp\left(-\frac{D n^2 \pi^2t}{(a+b)^2}\right)
\end{equation}
We then derive the Laplace transform of each distribution, defined as $\tilde{P}_s(s)=\int_0^\infty dt\,e^{-st}P_s(t)$ and similarly for $Q_0$, as 
\begin{equation}
    \tilde{P}_s(s) = \sum_{n=1}^\infty\frac{2}{\pi n}  \sin \left(\frac{\pi  a n}{a+b}\right) \left(\cos \left(\frac{\pi  a n}{2 (a+b)}\right)-\cos \left(\frac{\pi  n (2 a+b)}{2 (a+b)}\right)\right) \left( \frac{Dn^2 \pi^2}{(a+b)^2}+s\right)^{-1}
\end{equation}
and
\begin{equation}
    \tilde{Q}_0(s) = \sum_{n=1}^\infty  \frac{2 (1-(-1)^n)}{\pi n} \sin\left(\frac{\pi  a n}{a+b}\right) \left( \frac{Dn^2 \pi^2}{(a+b)^2}+s\right)^{-1}
\end{equation}
These expressions give us an analytic form for the first passage time distribution in Laplace space through Eq.\,\eqref{eq:SI_fptd} and the mean first-passage time through Eq.\,\eqref{eq:SI_mfpt}. For the case where $a=b=\ell$, we can evaluate the infinite sums analytically to write
\begin{equation}
    \tilde{P}_s(s) = \frac{1-\cosh \left(\sqrt{\frac{s \ell^2}{4D}}\right) \sech\left(\sqrt{\frac{s \ell^2}{D}}\right)}{s}
\end{equation}
and
\begin{equation}
    \tilde{Q}_0(s) = \frac{2}{s}\sech\left(\sqrt{\frac{s \ell^2}{D}}\right)\sinh^2\left(\sqrt{\frac{s \ell^2}{4D}}\right) = \frac{1}{s}\left[1 - \sech\left( \sqrt{\frac{s \ell^2}{D}}\right) \right],
\end{equation}
leading to 
\begin{equation}
\tilde{Q}(s) =\frac{2\sinh^2\bigg(\sqrt{\frac{(s+k_r)l^2}{4 D}}\bigg)}{k_r\cosh\bigg(\sqrt{\frac{(s+k_r)l^2}{4 D}}\bigg)+ s\cosh\bigg(2\sqrt{\frac{(s+k_r)l^2}{4 D}}\bigg)}.
\end{equation}

From this, we can evaluate the first-passage time distribution in Laplace space as
\begin{equation}
\tilde{\tau}(s) \equiv 1-s\tilde{Q}(s) = \frac{s + k_r \cosh\bigg(\sqrt{\frac{(s+k_r)l^2}{4 D}}\bigg)}{k_r\cosh\bigg(\sqrt{\frac{(s+k_r)l^2}{4 D}}\bigg)+ s\cosh\bigg(2\sqrt{\frac{(s+k_r)l^2}{4 D}}\bigg)}.
\end{equation}
The mean first-passage time is then given by
\begin{equation}\label{eq:alpha}
    \langle \tau\rangle \equiv \tilde{Q}(s=0) = \frac{2}{k_r}\sinh(\alpha)\tanh(\alpha)
\end{equation}
where we have defined $\alpha = \sqrt{\frac{k_r \ell^2}{4D}}$.

\subsection{Splitting probabilities}

Now we evaluate the fraction of trajectories that reach the left boundary compared to the right boundary, i.e. the splitting probability on the interval $(-a, b)$. We do so by first identifying the relevant events for evaluating the probabilities: for a particle initially at $0$, there are 5 outcomes that we identify for the end of the particles trajectory in a single resetting cycle: either the particle is reset to one of the three sites (at $-a$, $0$ or $b$) or the particle will diffuse through one of the boundaries at $-a$ or $b$ before any resetting event occurs.  

We can evaluate each of these 5 probabilities from the function $u(x, t)$ derived above. At the next resetting event, the probabilities for being reset to each of the three relevant sites take the form
\begin{gather}
    P(0\mapsto -a) = \int_0^\infty dt \int_{-a}^{-a/2}dx\, {u}(x, t)k_r e^{-k_r t} = \sum_{n=1}^\infty \frac{4 k_r \sin ^2\left(\frac{\pi  a n}{4 (a+b)}\right) \sin \left(\frac{\pi  a n}{a+b}\right)}{\pi  n \left(\frac{\pi ^2 D n^2}{(a+b)^2}+k_r\right)}\\
    P(0\mapsto 0) = \int_0^\infty dt \int_{-a/2}^{b/2}dx\, {u}(x, t)k_r e^{-k_r t} = \sum_{n=1}^\infty \frac{2 k_r \sin \left(\frac{\pi  a n}{a+b}\right) \left(\cos \left(\frac{\pi  a n}{2 (a+b)}\right)-\cos \left(\frac{\pi  n (2 a+b)}{2 (a+b)}\right)\right)}{\frac{\pi ^3 D n^3}{(a+b)^2}+\pi  k_r n}\\
    P(0 \mapsto b) = \int_0^\infty dt \int_{b/2}^{b}dx\, {u}(x, t)k_r e^{-k_r t} = \sum_{n=1}^\infty\frac{2 k_r \sin \left(\frac{\pi  a n}{a+b}\right) \left(\cos \left(\frac{\pi  n (2 a+b)}{2 (a+b)}\right)-(-1)^n\right)}{\frac{\pi ^3D n^3}{(a+b)^2}+\pi  k_r n}.
\end{gather}

If the particle is not reset to one of the sites, it has reached one of the two absorbing boundaries by pure diffusion. To evaluate the probability of this event happening, we first evaluate the time-dependent probability flux through each boundary: 
\begin{equation}
    \text{Left:}\quad J_{-a}(t) = D\partial_x {u}(x,t)\bigg|_{x=-a}\quad \text{and}\quad \text{Right:}\quad J_b(t) = -D\partial_x {u}(x,t)\bigg|_{x=b}.
\end{equation}
The probability of diffusion through the left boundary before being reset is the time integral of these boundary fluxes:
\begin{equation}
    \text{Left:}\quad\mathbb{P}(0\leadsto -a) = \int_0^{\infty}dt k_r e^{-k_rt}\int_0^tdt'\,J_{-a}(t') = \sum_{n=1}^\infty \frac{2 \pi  D n \sin \left(\frac{\pi  a n}{a+b}\right)}{\pi ^2 D n^2+k_r(a+b)^2}
\end{equation}
and
\begin{equation}\text{Right:}\quad  \mathbb{P}(0\leadsto b) = \int_0^{\infty}dt k_r e^{-k_rt}\int_0^tdt'\,J_b(t') = \sum_{n=1}^\infty\frac{2 \pi  D n (-1)^{n-1} \sin \left(\frac{\pi  a n}{a+b}\right)}{k_r (a+b)^2+\pi ^2 D n^2}.
\end{equation}
The splitting probabilities can then be evaluated from the one step probabilities as
\begin{subequations}\label{eq:split}
\begin{align}
    \mathbb{P}(0\rightarrow -a) &= \frac{\mathbb{P}(0\leadsto -a) + \mathbb{P}(0\mapsto -a)}{\sum_{s=-a, b}\mathbb{P}(0\leadsto s) + \mathbb{P}(0\mapsto s)}, \\
    \mathbb{P}(0\rightarrow b)  &= \frac{\mathbb{P}(0\leadsto b) + \mathbb{P}(0\mapsto b)}{\sum_{s=-a, b}\mathbb{P}(0\leadsto s) + \mathbb{P}(0\mapsto s)}.
\end{align}
\end{subequations}
Notice that the probability of being reset to the initial site $0$ does not feature here. To see this, note that the \textit{total} probability for, say, the particle to be absorbed by pure diffusion to the left boundary, is not equal to $\mathbb{P}(0\leadsto -a)$, but actually takes the form 
\begin{equation}
    \mathbb{P}(\text{Particle reaches }(-a)\text{ by pure diffusion}) = \mathbb{P}(0\leadsto -a) \sum_{m=0}^\infty (P(0\mapsto 0))^m
\end{equation}
as the particle may be reset to the initial site before reaching either boundary, effectively restarting the process. The other three outcomes (two at each boundary) can be written in a similar form and upon writing the fraction of the form in Eq.\,\eqref{eq:split}, the contributions from $P(0\mapsto 0)$ cancel, leading to Eq.\,\eqref{eq:split}.

\section{Details of numerical simulations of the microscopic model}

To simulate the microscopic dynamics of a single particle following the nearest-neighbor resetting process presented in the main text, we proceed as follows: (1) we determine when the next resetting event occurs by sampling a waiting time from an exponential distribution with rate $k_r$, (2) in between resetting events, we solve a purely diffusive Brownian dynamics using the Euler-Maruyama method with timestep $dt=10^{-5}$, (3) when the time for a resetting event is reached in the simulation, the position of the particle is reset to the nearest resetting site and a new waiting time is drawn from the exponential distribution. The total trajectory length was set to $T=1000$ in simulations.

To measure the effective diffusion coefficient, we record the mean-squared displacements of the particle position, defined as 
\begin{equation}
    \langle \Delta^2 x \rangle (\tau)= \frac{dt}{T-\tau} \sum_{t=0}^{T-\tau} (X(t+\tau)-X(t))^2.
\end{equation}
We then average this over $10^3$ realizations of the process, eventually evaluating the diffusion coefficient as the linear fit to the mean-squared displacement at large times, following the usual definition for 1D diffusive processes:
\begin{equation}
    D_{\rm eff} = \lim_{t\rightarrow \infty} \frac{\langle \Delta ^2 x\rangle(t)}{2t}. 
\end{equation}

\section{Any array of resetting sites can enhance self-diffusion}

In the main text, we derive an exact solution for the macroscopic diffusion coefficient of a diffusive particle under the nearest-neighbor resetting mechanism for a regular array of resetting sites. This is tractable as the stationary probability $\bm{\pi}$ is trivial for a regular array due to the symmetry of the problem and one obtains $\pi_n\equiv 1/N$. For arbitrary arrays of sites, this is no longer trivial and the calculation of the stationary probability requires us to solve $N$ fully-coupled linear equations which in general is not possible analytically. Here, we show that any array of resetting sites can enhance diffusivity when resetting happens sufficiently rarely. We do so by constructing a lower bound for the rescaled effective diffusion coefficient in the limit of small $k_r$. 

\subsection{Mapping to the Unit Interval}

For any given array of sites $\{x_i\}_{i\in[1,N]}$, we have shown in the main text that we can write the rescaled effective diffusion coefficient as a weighted sum of the local diffusion coefficients for each interval:
\begin{equation}\label{eq:si_sum_Deff}
    \frac{D_{\rm eff}}{D}= \sum_{i=1}^N {\pi_i}\left(\frac{D^{(i)}_{\rm eff}}{D}\right).
\end{equation}
The local diffusion coefficients for each interval $\{x_{i-1}, x_i, x_{i+1}\}$ can be mapped trivially to those on the interval $\{-a_i, 0, b_i\}$ by translation, i.e. for a suitable choice of $a_i$ and $b_i$. We now claim that the local diffusion coefficient $D_{\rm eff}^{(i)}/D$ remains invariant under the simultaneous re-scaling of the interval size and the diffusion coefficient $D$. In particular, we define the lengthscale $\ell_i = (a_i+b_i)/2$ and simultaneously rescale the interval $\{-a_i, 0, b_i\}$ to $\{-a_i/2\ell_i, 0, b_i/2\ell_i\}$ and the diffusion coefficient $D \rightarrow D/4\ell_i^2$. After this re-scaling, the mean first-passage time to adjacent sites remains the same, but the change in the average increase in the variance is rescaled as $\langle (\Delta x)^2\rangle_i\rightarrow \langle (\Delta x)^2\rangle_i / 4\ell^2_i $.  Overall, this means the rescaled local effective diffusion coefficient remains the same:
\begin{equation}
    \frac{\langle (\Delta x)^2\rangle_i}{2D \langle\tau\rangle_i} \longrightarrow \frac{\langle (\Delta x)^2\rangle_i/ (4\ell_i^2)}{2(D/4\ell_i^2) \langle\tau\rangle_i} \equiv \frac{\langle (\Delta x)^2\rangle_i}{2D \langle\tau\rangle_i}.
\end{equation}
Finally, the local diffusion coefficient for the interval $\{-a_i, 0, b_i\}/2\ell_i$ is the same as the interval $\{0, y_i = a_i/(a_i+b_i), 1\}$ (the two are equivalent up to translation), thus we have reduced our problem of finding a lower bound for the $N$ intervals $\{x_{i-1}, x_i, x_{i+1}\}$ to finding a lower bound for $N$ intervals of the form $\{0, y_i, 1\}$, each with the (interval-dependent) diffusion coefficient $D_i = D/4\ell_i^2$ and resetting rate $k_r$.

\begin{figure}[b!]
    \centering
    \includegraphics[width=0.4\textwidth]{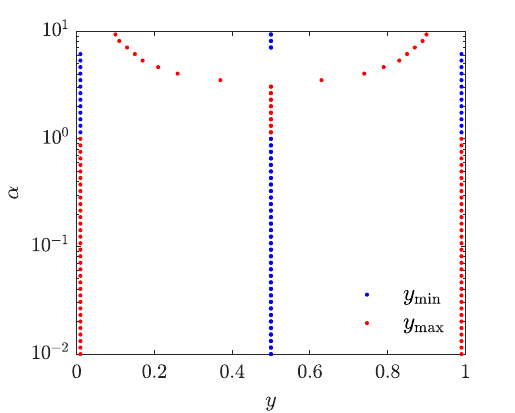}
    \caption{\textit{Location of the site $y$ to observe maximal and minimal effective diffusivity ---} The site placement that realizes a minimum (resp. maximum) of the effective diffusivity, denoted $y_{\rm min}$ (resp. $y_{\rm max}$), is plotted as a function of $\alpha$. For small $\alpha$, we observe that $y_{\rm min}=0.5$, while $y=0.5$ becomes a maximum of the effective diffusivity for moderate values of $\alpha$. We use this to analytically derive the behavior of the diffusivity to lowest-order in $\alpha$ for low values of $\alpha$ as given in Eq.\,\eqref{eq:pert_alpha}.}
    \label{fig:Deffmin}
\end{figure}

\subsection{Forming Lower Bound from Results for Unit Interval}

For the unit interval, we define the function $\mathcal{D}(y, \alpha)$ which is exactly the local diffusion coefficient for the interval $\{0, y, 1\}$ for $\alpha = \sqrt{k_r/16D}$, consistent with the initial definition of $\alpha$ in Eq.\,\eqref{eq:alpha}: there we defined $\alpha = \sqrt{k_r \ell^2 / 4 D}$ but now $\ell=1/2$. Indeed, the local diffusion coefficient for this interval is invariant under a re-scaling of $k_r$ and $D$ that keeps this $\alpha$ constant, which can be demonstrated through a formal nondimensionalisation of the problem.

We can bound the local diffusion coefficients trivially by writing
\begin{equation}
    \frac{D^{(i)}_{\rm eff}}{D}  = \mathcal{D}(y_i, \alpha_i) \geq \min_y \mathcal{D}(y, \alpha_i).
\end{equation}
In turn, this allows us to define our lower bound for the macroscopic diffusion coefficient in terms of our results for the unit interval in the form
\begin{equation}\label{eq:bound}
    \frac{D_{\rm eff}}{D} \geq\min_{i\in[1, N]}\left(\frac{D^{(i)}_{\rm eff}}{D}\right) \geq \min_{\{\alpha_i\}} \left(\min_y \mathcal{D}(y, \alpha_i)\right)
\end{equation}

We now look to evaluate the right-hand side of Eq.\,\eqref{eq:bound}. 
We note that this quantity is always positive for small values of $\alpha$, implying that diffusion is boosted for any value of $y$ when $k_r$ is sufficiently small. 

In Fig.\,\ref{fig:Deffmin}, we plot the values of $y$ at which the minima and maxima of the effective diffusivity are found; in particular, for $\alpha\ll1$, the minimum occurs at $y=0.5$. For this value of $y$, we have previously shown that the local diffusion coefficient for the unit interval can be computed exactly:
\begin{equation}
    \mathcal{D}(y=0.5, \alpha) = \alpha^2 \csch(\alpha)\coth(\alpha).
\end{equation}
We now expand this result to leading-order in $\alpha$ (treating it as a small parameter) to derive
\begin{equation}\label{eq:pert_alpha}
 \min_y\left[\mathcal{D}(y, \alpha)-1\right] \geq  \frac{\alpha^2}{6} + \mathcal{O}(\alpha^4) > 0.
\end{equation}

Finally, this gives us the lower bound for the macroscopic diffusion coefficient: suppose that $k_r$ is sufficiently small such that $\alpha_i\ll 1$ for each of the $N$ intervals, then we have the bound 
\begin{equation}
    D_{\rm eff}/D \geq \min_{\alpha_i} \left[1 +  \frac{\alpha_i^2}{6} + \mathcal{O}(\alpha_i^4) \right] > 1,
\end{equation}
thus demonstrating that the effective macroscopic diffusion coefficient can be boosted for any arbitrary array of sites.

\section{Lattice is optimal for enhancing diffusion: Numerical analysis}

\begin{figure}[b!]
    \centering
    \includegraphics[width=\textwidth]{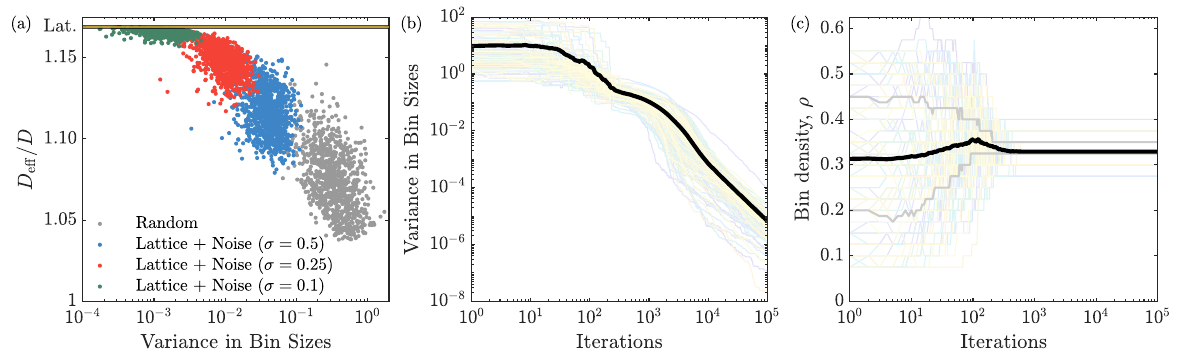}
    \caption{\textit{A regular array optimizes the enhancement of self-diffusion ---} (a) Comparing the maximal enhancement in diffusion for the regular array of resetting sites to randomly generated arrays, we see that the lattice always outperforms others. There is also a correlation between variance in distances between sites and maximal diffusivity. (b) We design an evolutionary algorithm to determine arrays of sites that locally maximize the increase in effective diffusivity (see Sec.\,\ref{sec:evoalg}). We observe that such sites necessarily have lattice structure (regular spacing between sites) as the variance always decreases with the number of iterations. (c) Finally, we set $D=k_r=1$ and allow sites to be added and removed randomly at each step of the process. In this case, we observe that, on average, the total density of sites $\rho\rightarrow \sqrt{\frac{1}{4(\alpha^*)^2}}\approx0.3113$ where $\alpha^* = 1.606$.}
    \label{fig:numerical_comp}
\end{figure}

\subsection{Lattice outperforms randomly generated arrays} 
Given that all arrays can enhance diffusion for some value of the resetting rate, we now consider the optimal arrays for enhancing the diffusion coefficient. In the main text, we showed explicitly that the maximum possible enhancement is obtained for regular arrays of resetting sites. We confirm this here by comparing the diffusion-enhancement of the lattice to the maximum diffusivity obtainable for other arrays of sites. We first generate arrays where sites are uniformly distributed between $0$ and $L$, then evaluate the maximum value that the effective diffusion coefficient can take numerically by varying the resetting rate $k_r$. These results are shown in grey in Fig.\,\ref{fig:numerical_comp}(a), falling always below the maximum value of the diffusivity for regular arrays, indicated by the solid line $D_{\rm eff}/D\approx1.169$. 

It appears from these results that the arrays with less variance in the distances between neighboring sites perform the best. A regular array of evenly-spaced resetting sites has fixed spacing between sites and hence trivially zero variance. To illustrate this, we generate perturbations of the regular array, with sites displaced from where they would be in a lattice arrangement by some random amount drawn from a zero-mean Gaussian distribution with standard deviation $\sigma$. We see that the arrays with smallest $\sigma$ perform best, in agreement with the idea that order enhances diffusivity for our nearest-neighbor resetting mechanism.

\subsection{Evolutionary algorithm to (locally) maximize the boost in diffusivity} \label{sec:evoalg}
Finally, we construct an evolutionary algorithm to find arrays of resetting sites that (locally) maximize the diffusion coefficient in the full $N$-dimensional space of resetting arrays, $\{x_0, \dots, x_{N-1}\} \in S$, where $S$ is the subset of $\{0\}\times (0, L) ^{N-1}$ where $x_1<x_2<\dots <x_{N-1}$. We begin by generating $N$ uniformly distributed points and evaluating the maximal diffusion coefficient for this array by varying $k_r$ as above. We then make a small adjustment to the position of one of the resetting sites, leaving $x_0=0$ without loss of generality. If this new array exhibits a maximal effective diffusion coefficient that is greater than that of the previous array, we keep this change, otherwise it is discarded. 

After many iterations, we observe that the long-time arrays always exhibit a regular structure. To quantify this, we measure the variance in the distances between sites $d_i = |x_{i+1}-x_i|$ as a function of time and average over many realizations, as shown in Fig.\,\ref{fig:numerical_comp}(b). We observe that this quantity monotonically decreases at large times, which implies that the array of sites that the algorithm is converging towards has a lattice structure.

We then consider the case where the number of sites can fluctuate, instead fixing the resetting rate $k_r$. We implement this by saying that at each iteration, the algorithm can randomly move a site (as above) or add a site placed randomly in $(0, L)$ or remove a randomly-chosen site (except the site at $x_0=0$), each with probability $1/3$. In this case, we similarly observe a convergence of the variance in the inter-site distances towards zero, but also the resetting sites density converges towards a common value, which we identify as the density $\rho$ satisfying
\begin{equation}
    \rho\rightarrow \sqrt{\frac{k_r}{4(\alpha^*)^2D}}
\end{equation}
where $\alpha^* = 1.606\dots$ [see Fig.\,\ref{fig:numerical_comp}(c)].


%